\documentstyle[draft,psfig]{mn}
\def\zl{{z_{\rm lens}}}
\def\zr{{z_{\rm gal}}}
\def\gs{\mathrel{\raise0.35ex\hbox{$\scriptstyle >$}\kern-0.6em 
\lower0.40ex\hbox{{$\scriptstyle \sim$}}}}
\def\ls{\mathrel{\raise0.35ex\hbox{$\scriptstyle <$}\kern-0.6em 
\lower0.40ex\hbox{{$\scriptstyle \sim$}}}}

\title[Weak Lensing in the Fields of 3C Sources]
      {A Weak Lensing Survey in the Fields of z$\sim$1 Luminous Radio Sources}
\author[Bower \& Smail]{ 
Richard G.\ Bower \& Ian Smail\\
Department of Physics, University of Durham, South Rd., Durham DH1 3LE.
}
\begin{document}

\date{
 Received -- ---- 1996; in original form 1996 Dec 18}

\label{firstpage}

\maketitle

\begin{abstract}
In this paper we present weak lensing observations of the fields around
8 $z\sim 1$ luminous radio sources. These data are used to search for
the lensing signatures of galaxy clusters that are either physically
associated with the radio objects, or are foreground systems projected
along the line of sight. The radio sources were all the subjects of
deep imaging with WFPC-2 on the {\it Hubble Space Telescope} providing
high quality shape information on large numbers of faint galaxies
around these sources.  Statistical analysis of the coherent shear field
visible in the shapes of the faint galaxies indicates that we have
detected a weak lensing signal close to one of the targets, 3C336 at
$z=0.927$, with a high level of confidence.  A second, independent
WFPC-2 observation of this target reinforces this detection.  Our
results support the earlier suggestion of weak lensing in this field by
Fort et al.\ (1996) using ground-based data.  We also combined the
shear distributions in the remaining 7 field to improve our sensitivity
to weak shear signals from any structure typically associated with
these sources.  We find no detectable signal and estimate an upper
limit on the maximum shear allowed by our observations.   We estimate
likely limits for the redshift distribution of the faint galaxies used
in our analysis, building upon results from inversion redshift surveys
through moderate redshift cluster lenses, and the analysis of the
colour distributions of galaxies in the Hubble Deep Field.  Using
these, we convert our observed lensing signal and limits into estimates
of the masses of the various structures.  We suggest that further
lensing observations of distant radio sources and their host
environments may allow the cluster $L_X$--mass relationship to be
mapped at high-$z$.  This is crucial for interpreting the results of
the next generation of deep X-ray surveys, and thus for providing a
strong observational constraint on the redshift evolution of the
cluster mass function out to $z=1$.
\end{abstract}


\section{INTRODUCTION}

The properties of massive, collapsed structures at high redshift
provide a vital test of competing theories of structure formation in
the universe (e.g.\ Eke et al.\ 1996).  In particular, the high masses
associated with the richest clusters place them on the extreme tail of
collapsed structures at their epochs, this in turn makes their space
density an extremely sensitive test of models of cosmological structure
formation.  For example, by indicating a slow pace of evolution, an
extensive population of {\it massive} clusters at $z\sim1$ would
present an impossible challenge to cosmological models with high
density (i.e.\ $\Omega_o \sim 1$).

It is for this reason that a number of groups have undertaken surveys
for distant clusters, using projected galaxy density (e.g.\ Abell 1958;
Gunn, Hoessel \& Oke 1986; Couch et al.\ 1991; Postman et al.\ 1996) or
X-ray emission from the gas bound to the cluster potential (e.g.\ Henry
et al.\ 1992; Castander et al.\ 1995; Rosati et al.\ 1995; Bower et
al.\ 1996).  While the possible pit-falls associated with optical
searches are well documented (c.f.\ Couch et al.\ 1991), the problems
with interpreting the X-ray results are only now being explored (see
below).  To an extent these searches have succeeded: a number of
optically selected $z\sim 1$ clusters are known, while the highest
redshift X-ray selected clusters are at $z=0.83$ and $z=0.78$ (Gioia \&
Luppino 1994).  However, the total number of confirmed clusters at
$z\sim 1$ still only amounts to a handful and this seriously limits
studies of the properties of these systems and their constituent
galaxies.
  
While the next generation of X-ray surveys will provide more candidates
for $z\sim 1$ clusters, there are already indications that these will
not be the salvation we might hope.  Firstly, the number of such
distant systems detected to date is far below that expected by simple
models for hierarchical growth in the universe.  This implies that the
evolution of the X-ray luminosity function is more complex than was
first anticipated.  It appears that the changes in the X-ray emission
from distant clusters arise not only from the growth of the cluster
potential, but also from the redshift dependence of the concentration
of gas within the cluster (Evrard \& Henry 1991; Bower 1996). We are
thus left with the possibility that the observed X-ray evolution is
driven both by details of the emission within the clusters, and the
underlying growth of the cluster. To unambiguously interpret the X-ray
surveys in terms of evolution in the cluster masses, we need to
calibrate the X-ray properties of distant clusters with other measures
of the cluster mass across a range of epochs. Only by combining these
two pieces of information can we determine how the observed space
density of distant clusters constrains cosmogonic models of structure
formation.  The most direct mass estimate to use in this comparison is
the gravitational lensing signature of the cluster.

Gravitational lensing has become one of the most widely applied tools
to study the mass distributions in moderate redshift clusters (see Fort
\& Mellier 1994 for a recent review).  The strength and direction of
the coherent distortions induced in the shapes of {\it all} background
field galaxies seen through the cluster contains information about both
the distribution and amount of mass in the cluster (Tyson, Valdes \&
Wenk 1990; Smail et al.\ 1995a), as well as the relative distances of
the cluster lens and the distant field population (Kneib et al.\ 1994;
Smail, Ellis \& Fitchett 1994).  Indeed, the obvious next step of just
using lensing to select clusters is already underway with the new
generation of panoramic imagers on large ground-based telescopes.
Lensing analysis of the shear patterns in background field galaxies
will provide a mass selected catalogue of clusters and groups at $z\ls
0.5$.  Unfortunately, extending this approach to $z\sim 1$ requires
measuring the shear signal at very faint magnitudes and small angular
sizes for the distant field galaxies, which is difficult from the
ground.

As a first step in comparing X-ray and lensing mass estimates for a
large sample of distant clusters, Smail et al.\ (1996a) have analysed
the gravitational lensing signal detected in deep {\it Hubble Space
Telescope} (HST) observations using the Wide Field and Planetary Camera
(WFPC-2) of 12 rich clusters covering a wide range of X-ray
luminosities.  The mean redshift of this sample is $z\sim 0.4$ and they
detect shear signals in the fields of 90\% of the clusters.   As they
show, the strength of these shear measurements correlates well with the
X-ray luminosity of the clusters.  Moreover, by assuming a redshift
distribution for the faint field galaxies, in agreement with that
observed to $R=25$ by Kneib et al.\ (1996), Smail et al.\ converted
these shear measurements into mass estimates and compared these to the
values inferred from the X-ray luminosities.  The observed correlation
implies at most a modest decrease in the cluster mass at a fixed
luminosity out to $z=0.4$.  Smail et al.\ (1996a) stress the
observations are compatible with {\bf no} evolution in the X-ray
luminosity versus mass ($L_X$--$M$) correlation from $z=0$ to $z=0.4$
at the 99\% level.

Bower (1996) discusses the interplay between the evolution of the
$L_X$--$M$ correlation and the cluster X-ray luminosity function.
Reviewing the limits  set at $z\sim 0.4$ shows that even an approximate
measurement of the $L_X$--$M$ relation at $z\sim 1$ would, combined
with the results of the current generation of deep X-ray surveys,
provide a strong observational constraint on the redshift evolution of
the cluster mass function out to $z=1$. 
  
In the known higher redshift clusters, the comparison of X-ray and
lensing masses has only been undertaken for two systems, 3C324 at
$z=1.24$ (Smail \& Dickinson 1995) and MS1054$-$03 at $z=0.83$ (Luppino
\& Kaiser 1996).  Using a 64~ks HST/WFPC2 exposure of 3C324, Smail \&
Dickinson (1995) detected the coherent gravitational distortion created
in the images of background field galaxies by the cluster potential.
The average shear amplitude was 3\% within a 250 $h^{-1}$ kpc radius
aperture centered close to the radio source.   By assuming a maximum
average redshift ($<\! z\! >=2$) for the faint field population at the
magnitude limit of their exposure, supported by estimates from detailed
modelling of strong cluster lenses at $z\sim 0.2$ (Kneib et al.\ 1996)
they converted this detection into a lower limit on the cluster mass
within the same 250$h^{-1}$ kpc aperture: $M_{250} \geq 1.2\times
10^{14} \,h^{-1}M_{\odot}$.  For comparison, the cluster's X-ray
luminosity is $L_X($0.3-3.5$) = 1.3 \times 10^{44} h^{-2}$ erg/sec,
indicating a comparable mass normalisation of the $L_X$--$M$
correlation to that observed in the local universe.  Luppino \& Kaiser
(1996) also find rough agreement between their X-ray and lensing masses
within the framework of the local $M$--$L_X$ relation.  These two
luminous X-ray clusters thus give some indication of the $M$--$L_X$
relation at $z\sim 1$. However to robustly map the $M$--$L_X$ relation
at $z\sim 1$  we require lensing observations of clusters across a
much wider range of X-ray luminosity.

The success of these observations has prompted us to approach the
problem from a different angle. At intermediate redshifts ($z\sim0.5$),
radio-loud quasars have been shown to occupy high density regions
(Ellingson et al.\ 1991), specifically they appear to reside in the
central regions of moderate richness clusters.    The presence of
radio-loud quasars within the cores of their host clusters is discussed
by Yee \& Ellingson (1993), who associate this with the formation of
QSOs due to the high gas and galaxy densities in these environments.
This result has been extended to higher redshifts (Hintzen et
al.\ 1991; Hutchings et al.\ 1993, 1995) on the basis of deep
ground-based observations using broad- and narrow-band imaging.   At
even higher redshifts, examples are known of radio sources inhabiting
high density environments (e.g.\ Pascarelle et al.\ 1996).  Our aim is
therefore to use radio sources as tracers of high density environments
in the distant universe. Deep lensing observations of the fields around
these objects will be sensitive to mass concentrations associated with
the radio sources.  In this way we hope to expand the sample of distant
clusters known, as well as providing the lensing mass estimate
necessary to determine the $M$--$L_X$ relation for these distant
clusters.  The advantage of this approach over one starting from a
flux-limited X-ray catalog are the much wider range of X-ray
luminosities spanned by the final sample (vis.\ $L_X \propto M^3$).

Furthermore, a sample of confirmed massive clusters at $z\sim 1$ would
also have considerable interest for studies in a number of other
areas.  As shown by Smail, Ellis \& Fitchett (1994), a massive cluster
at $z\sim 1$ also has tremendous potential for furthering our
understanding of the redshift distribution of the very faint field
population.  Moreover, an investigation of the galaxy populations
within such clusters is well suited to uncovering observational evidence
for environmentally-driven galaxy evolution (Smail et al.\ 1996b).

While the lensing analysis is a sensitive probe of mass concentrations
associated with the distant radio sources, it will also detect
foreground mass concentrations along the line of sight.  If these are
sufficiently massive their shear fields may resemble a rich cluster at
the radio source redshift.  Although such structures would not be
expected to contribute a significant shear signals for any random
sightline, these fields may not necessarily be truly `random', in so
much as they contain distant, luminous radio sources.  At issue is
whether amplification by any foreground mass structure has also
affected the distant radio source, causing it to be included in a
catalog of bright radio sources.  This has been a subject of much
discussion, especially in regards to the 3C catalogue (Hammer et
al.\ 1986; Hammer \& Le F\`evre 1990), where the steep source counts
are claimed to make amplification bias a particular problem.  Most of
the studies of the possible effects of amplification bias in the 3C
catalog have relied upon identification of the putative foreground lens
(usually expected to be a single galaxy halo -- owing to their much
large lensing cross section) using direct imaging. Only recently have
studies started to search for more massive foreground structures by
identifying the weak shear of the background field population produced
by the same lens which may be also amplifing the radio source (Bonnet
et al.\ 1993; Fort et al.\ 1996).  However, these searches have so far
produced only inconclusive results.  The linked issues of the role of
amplification bias on distant radio sources and the existence of mass
structures associated with the radio sources are therefore still of
considerable interest.

In this paper we report the analysis of archival HST/WFPC-2
observations of powerful radio sources at $z\sim1$. The data allow us
to search for the signature of gravitational lensing from any
associated galaxy clusters.  The use of HST provides a high sensitivity
for the detection of weak shear, far superior to that obtainable in
even the best conditions with ground-based telescopes. Its main
drawback is the restricted field of view (although similarly small
fields have been used in most of the ground-based studies to date)
which limits our detection of mass concentrations around the distant
radio sources to those systems where the radio source lies within $\sim
500 h^{-1}$ kpc of the cluster core.  As we have discussed above this
does not appear to be  a restriction for the distant radio sources,
where a large fraction are expected to lie close to the centres of any
structures (c.f.\ Hutchings et al.\ 1995), an assumption we can test.
The small field will also not produce a significant bias against
detecting any foreground structures which are amplifying the distant
radio sources, as these are only likely to provide significant
amplification if the radio source lies behind the core regions of the
foreground lens.  Nevertheless, our primary goal is to build up a
catalogue of clusters at high redshift which are confirmed as massive
structures through the strength of their lensing signal.  We present in
this work on our first results on the shear fields in a sample of 8
radio galaxies (RG) and radio-loud QSOs (RLQ) at $z\sim 1$.  We begin
in \S 2 by detailing the sample, the observations and their reduction.
\S 3 presents our analysis of the shapes of the faint field galaxies
seen in our exposures.  We also give the results on the limits on the
coherent shear seen in these fields in  \S 4.  In \S 5 translate our
shear measurements into mass estimates for structures in the fields and
discuss these limits, before giving our main conclusions in \S 6.

\section{REDUCTION}

The data analysed in this paper were all taken with the WFPC-2 onboard
the HST.  These data were retrieved from the HST Archive operated by
ST-ECF.  We have selected all $z\sim 1$ radio galaxies or QSOs which
have been imaged with WFPC-2 for $>5$ ks in any of the standard
wide-red filters and were available in the Archive in early-1996.  The
majority of these data come from an imaging survey undertaken by
Stockton \& Ridgway (GO\# 5401) of a complete sample of $z\sim1$ radio
galaxies and QSOs.  In addition, a second, deeper observation of one of
these targets, the QSO 3C336, was taken as part of GO\# 5304
(P.I.\ Steidel), this field is called 3C336\#2 to differentiate it from
the GO\# 5401 exposure.  The objects comprise a complete sample of
$z\sim1$ luminous radio galaxies and radio-loud QSOs in the 3C
catalogue (Laing, Riley \& Longair 1983), selected on the basis of
their extended radio flux density.    These targets were observed in
F622W, F675W or F702W.  We give a log of the observations in Table~1.

The shorter integrations comprise 4 orbits on each target, each orbit
is split into two exposures to allow removal of cosmic ray events,
these 4 orbits are themselves offset by $\pm 10.5$ WFPC-2 pixels to
allow hot pixel rejection.  The longer integrations consist of 4
exposures (2 orbits) at each pointing, rather than 1. The exception is
the GO-5304 exposure of 3C336\#2 which is in 3 groups of 4 orbits shifted
by integer pixel offsets, with each orbit split into 2 exposures.
After standard pipeline reduction, the exposures at each pointing were
combined using the {\sc IRAF/STSDAS} task {\sc CRREJ}, before being
aligned using sub-pixel shifts (when required) and high-order
polynomial interpolation, and finally coadded.  The PC chip was
removed, owing to its lower surface-brightness sensitivity and the sky
levels in the remaining 3 WFC chips equalized, before mosaicing them
together.  The mosaicing uses integer pixel shifts to roughly position
the 3 chips relative to each other.  This is sufficient to align
objects lying across the chip boundaries at the $\ls 1$ pixel level and
has the advantage over a true astrometric mosaic that the data is not
further resampled.  We chose to transform the various observed
passbands to the standard Cousins $R$ using the zero points and colour
corrections from Holtzman et al.\ (1995) and assuming a median color of
$(V-R)=0.6$ (Smail et al.\ 1995b).   The galactic extinction for these
radio sources given by {\sc NED} is at or below $A_R \sim 0.05$ and so
we have not applied any correction.  The final images cover a region of
$\sim 0.6 h^{-1}$ Mpc\footnote{We take $q_\circ = 0.5$ and $h = H_\circ/100$
km/sec/Mpc}  across at the radio source distance to the 50\%
completeness limits listed in Table~1.

To catalog faint objects in these frames and measure their shapes we
use the SExtractor image analysis package (Bertin \& Arnouts 1996).  We
adopt a detection isophote equivalent to $\sim 1.5 \sigma$ above the
sky, where $\sigma$ is the standard deviation of the sky noise measured
pixel to pixel and a minimum area after convolution with a 0.3 arcsec
diameter top-hat filter of 0.12 arcsec$^{2}$.  After automated
detection and deblending, the object catalogs for each frame were
visually inspected and spurious or unreliable objects were removed,
as well as those within 1.5 arcsec of any CCD boundary. In
several frames, a  proportion of the initial area was lost due to
contamination by stellar diffraction spikes.  The analysis of our
frames provides catalogs of 200--400 objects in each of the radio
galaxy fields.  

We apply three further cuts to the resulting catalogues before using
them in the lensing analysis.  The first two cuts are on magnitude.  We
discard all objects fainter than $R_{\rm max}$ (values for each frame
are given in Table~1) to remove objects with inadequate signal-to-noise
for measuring reliable shapes.  We also apply a bright magnitude cut to
reduce foreground contamination, retaining only galaxies with $R >
24.5$, the brightness of a non-evolving $L^*$ galaxy in a $z\sim 1$
cluster.   This should guarantee that a large fraction of the field
galaxies in our sample lie at redshifts $z\gs 1$.  Finally, we remove
all objects with ellipticities $\epsilon < 0.05$ or $\epsilon > 0.4$,
where $\epsilon = (a-b)/(a+b)$ and $a$ and $b$ are the major and minor
axis lengths of an object in order to exclude objects that are either
too round to accurately determine their orientations or are so
elongated that their ellipticity is almost certainly intrinsic (Mellier
\& Fort 1996). The final catalogues used in the lensing analysis
contain 100--170 objects in the shorter exposures, corresponding to
surface densities of 25--40 arcmin$^{-2}$ (Table~1). Significantly
higher object surface densities are obtained by stacking the data for
the individual radio sources as discussed in Section~4.2. 

\section{ANALYSIS}

\subsection{Shear Fields and Likelihood Maps}

To search for the effects of lensing we evaluate the complex shear:  $g
= \epsilon e^{2i\phi}$, where $\phi$ is the position angle of the major
axis of the object, for every object in the catalog (Kneib et
al.\ 1996).  In order to obtain an initial impression of the shear
field around each source, we binned the estimates from the individual
objects into independent $20'' \times 20''$ cells. The resulting
vectors can be plotted as a vector field over the image. An example of
this is shown in Figure~1.    

\begin{figure}
\begin{center}
\psfig{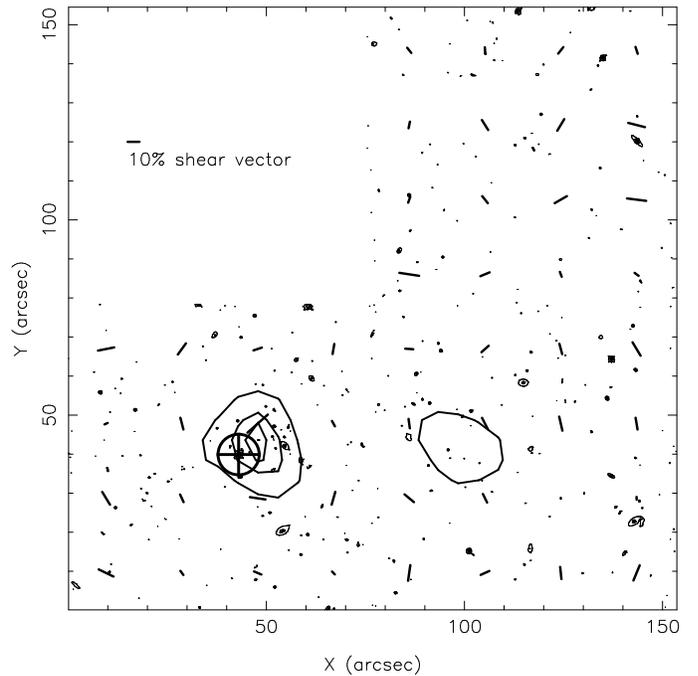}
\caption{An F702W image of the field around the radio source 3C336 at
$z=0.927$.  The field of view is $150 \times 150$ arcsec (0.63 $\times$
0.63 $h^{-1}$ Mpc).  Overlaid is a contour map of the shear alignment
probability from our lensing analysis.  The shear pattern, indicated by
the vector field, shows a strong coherence around the QSO position
(marked by the circled cross), with a peak likelihood of $- \log P =
3.30$. A less significant secondary peak lies to the right of the
primary peak. The contours show 80\%, 50\% and 10\% of the peak
probability.  A vector representing a 10\% observed shear is shown at
the upper left.
}
\end{center}
\end{figure}

In order to obtain a quantitative measurement of the shear field around
each of the sources, we adopted a smoothing technique based on Kaiser
\& Squires (1993, KS). At a fine grid of points overlayed on the image,
we constructed the probability distribution for the tangential
alignment of faint galaxies around that point. For this we calculate
the average tangential shear, $g_1 = - \epsilon \cos (2 \theta)$, where
$\theta$ is the angle between an object's major axis and the vector
joining it to the grid point, using a weighting function ($W(r)$) as
described by KS. The form of the fourier-space transfer function was
chosen to match the signal expected for an isothermal sphere, but with
high-frequency components suppressed according to the Weiner filter
criterion. If a cluster lies within, or close to, our frame, the shear
signal would be present (with falling amplitude) across the full extent
of the HST image: this is properly accounted for by the weighting
function we use.  The smoothing scale was chosen so as to give greatest
weight (ie.\ $W>0.8W_{\rm max}$) to galaxies lying in an annulus from
$r= 30$--$150 h^{-1}$ kpc around each point (Figure~2). This procedure
provides a quantitative measurement of the coherence of the shear
pattern around each point. In order to estimate its statistical
significance, we compare this to the distribution of this value for a
null shear field, estimated by bootstrap resampling the galaxy shapes.
The random catalogues are created by assigning each galaxy in the
original catalogue a new ellipticity and orientation drawn at random
from the other galaxies. This preserves the ellipticity distribution
and incomplete spatial coverage of the original catalogue but ensures
that any true lensing signal is eliminated. The shear signal in this
mock catalogue is then evaluated across the grid and the shear values
measured compared with those determined from the original catalogue.
Repeating the procedure 1000 times allows us to produce a map of the
significance of the observed signal. An example is shown as a contour
plot in Figure~1. The maps of the shear likelihood provide a useful
visually confirmation of the significance of any maxima associated with
a particular radio source. Finally, we note that while the filtering
scheme designed above is optimal for the detection of clusters with
isothermal mass profiles, we obtained almost identical results using
the Gaussian-filtered point mass algorithm described by KS.

\begin{figure}
\begin{center}
\psfig{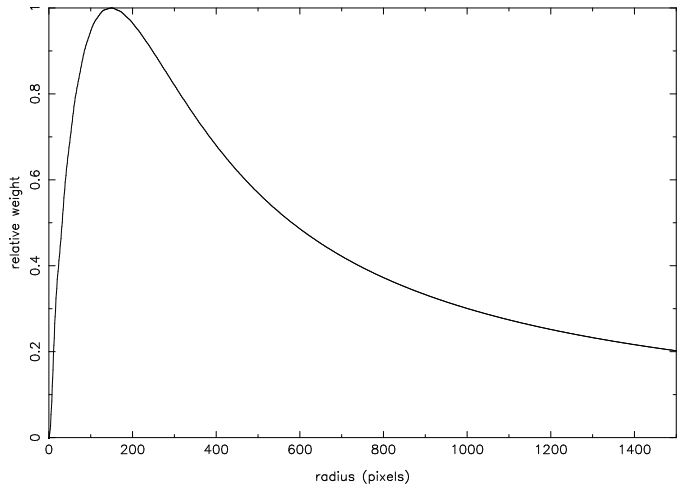}
\caption{The weighting function used in the calculation of mean 
shear strength. The function has been constructed by Weiner filtering
the fourier components of the shear signal expected for an iosthermal
mass distribution. The function has a slower radial fall-off than a
weighting function optimised for the detection of a point-mass 
concentration.}
\end{center}
\end{figure}

To determine the contribution from possible distortion in the
instrument optics we turn to the discussion in Smail \& Dickinson
(1995).  They use deep WFPC-2 pointings of blank fields from the Groth
Strip to estimate the possible instrumental shear signal at the
position of the detected mass concentration in their observations of
the field of 3C324.  Using the same deep, blank field observations we
estimate instrumental shear limits of $0.005\pm 0.006$\% (GO\# 5401)
and $0.008\pm 0.006$\% (GO\# 5304) for the positions of the target
radio sources in the present observations.  These are measured inside
a 300 $h^{-1}$ kpc diameter aperture centred on the relevant position
in the WFPC-2 field-of-view.

\subsection{Modelling}

The shear probability maps described above do not directly correspond to
the distribution of mass across the frame. Firstly, the spatial
coverage of the data is extremely limited. At the edges of the frame
the significance of a given lensing mass is suppressed because (i)~a
smaller number of objects contribute to the signal (after the weighting
scheme had been applied) thus enhancing the noise level, and (ii)~the
average shear signal is lower because of the greater distance of the
objects from the assumed centre of the mass concentration. Secondly,
our weighting scheme is based on the assumption that only one
isothermal mass concentration is present, and, for our choice of
weighting function, the resulting shear map does not correspond to a
simple convolution of the mass distribution. These difficulties are
extensively discussed by KS and Squires \& Kaiser (1995).  
 
In order to obtain an estimate of the mass associated with each of our
3C targets, we adopt a different approach that is based on simulating
the observed signal. Assuming a lens redshift and a redshift
distribution for the catalogued objects (see below), we calculate the
average shear signal expected as a function of the lensing cluster's
velocity dispersion.  To estimate the efficiency of our shear
measurements we use the same approach discussed by Smail et
al.\ (1996a), from this we estimate a mean shear measurement efficiency
of $0.8\pm 0.1$ for galaxy images with the signal-to-noise typical of
this study.  We include this factor in the calculation below.   The
resulting shear signal depends on the position of the mass within the
frame, and on the distribution of objects across the frame. A separate
calculation is therefore required for each of the catalogues listed in
Table~2. This establishes the procedure for estimating the lensing mass
from the mean shear signal as a fraction of the critical lensing mass.
The critical mass depends, however, on the redshift of the lens and the
redshift distribution ($N(z)$) of the faint galaxies used to map the
shear distortion.  Estimating the $N(z)$ at these depths is not a
straight forward task, and involves some uncertainty, although our
knowledge of the distances to the very faint field population is
rapidly improving (c.f.\ Kneib et al.\ 1996). We follow the procedure 
described by Smail \& Dickinson (1995), assuming that the redshift 
distribution has the form:
$$
N(z) \propto {z^2 e^{-(z/z_0)^\beta}}.
$$
For our fiducial model, we adopt $\beta=5$, for which $<\! z\! > = 0.78
z_0$.  where  $<\! z\! >$ is the mean redshift of the field
distribution. For our chosen magnitude limits the median redshift of
the field population is $<\! z\! > = 1.0\pm 0.2$ (Kneib et al.\ 1996)
for $R=24.5$--25.6,  this is close to the `no evolution' prediction and
hence we have used this model to extrapolate this $N(z)$ to slightly
fainter limits.  Using this model we obtain median redshifts of $<\!
z\! > \sim 1.1$ and $<\! z\! > \sim 1.2$ for magnitude limits of
$R=26.1$ and $R=27.0$ respectively.  Alternatively, using the galaxy
evolution models which fit the colour distributions in the HDF
(Metcalfe et al.\ 1996) we would expect $<\! z\! > \sim 2$.  We can now
use this $N(z)$ to estimate masses for any structure whose shear signal
we detect.   One interesting limit is to assume that the cluster is
physically associated with the radio source, and to adopt the
conservative assumption that only 50\% of the field population lie
beyond the lens, consistent with the $R\leq 25$ $N(z)$ claimed by Kneib
et al.\ (1996).  This yields an upper limit on the mass of the lens. If
a higher proportion of objects lie beyond the lens, either because the
lens is foreground to the radio source or because the median faint
object redshift is considerably greater than $z\sim 1$, the real mass
will be less.  For simplicity, the mass estimates we quote in the
following sections are given in terms of the velocity dispersion of the
isothermal halo. These can, however, be converted to the mass contained
within an $250\,h^{-1}\hbox{kpc}$ radius using the relation $M_{250}
\simeq 1.8\times10^{14} h^{-1} M_\odot (\sigma /
1000\,\hbox{km/s})^2$.

\section{RESULTS}

Table~2 sumarises the results of our shear analysis applied to each of
the 3C fields in turn. We restrict our selection to only those maxima
within a $200\,h^{-1}\hbox{kpc}$ diameter aperture around the radio
galaxy.   In only one case, 3C336\#2, do we detect a highly significant
($>3\sigma$) shear peak within this distance of the radio source.  We
discuss the interpretation of this signal in the next section.
Expanding our search to the entire WFPC-2 field, we find only one
additional example of a statistically significant peak ($-\log P \geq
3$), in 3C212.  Returning to the $200\,h^{-1}$ kpc detection aperture,
the next largest likelihood after 3C336 is $-\log P = 1.95$, measured
for 3C289.  Thus in 7 of the fields we find no evidence of a coherent
shear pattern to the limits of our data.   For these fields, we
estimate the 99\% confidence limit on our detections in order that we
can quantify the upper limit on any lensing mass. The detection limits
lie between 3.0 and~4.4\% (Table~2). We confirmed that these limits
were realistic by comparing these limits with the unconstrained maxima
seen in the shear significance maps.  Although no shear signal has been
detected in these images individually (3C336 and 3C336\#2 apart), we
may probe to deeper limits by co-adding the data-sets. The resulting
velocity dispersion limits from our observed shear fields are given in
Table~2, using the measured signal in the case of 3C336\#2, and the
detection limit in the other cases where no significant shear signal
was measured. 

\subsection{3C336 $z=0.927$}

In the case of 3C336, a highly significant ($- \log P = 3.30$) shear
signal is detected in the longer exposure F702W image (3C336\#2),
centred close to the radio source position.  It is possible with some
imagination to pick out a weak circular pattern in the shear vectors
around the radio source in Figure~1, although there are no good
candidates for strongly lensed features (arcs or pairs) in the field.
The amplitude of the average weighted shear is $4.0\pm 1.0$\%.  A
signal of similar amplitude is also seen in the shorter F622W image of
this field, close to the position of the shear peak in the F702W image,
confirming the detection.  However, the significance of this shear
pattern is lower due to the smaller number of objects in the F622W
catalogue.   It should be also noted that the position of the radio
source and the roll-angle of the two observations of 3C336 are
different, giving further confidence that the signal is not due to
distortions introduced by the WFPC-2 optics.  We have also used a
brighter galaxy sample, $R=22.0$--24.5, to test if the shear field is
detectable in these brighter galaxies (which are expected to
predominantly lie at $z\ls 1$).  We obtain an average shear of 2.3\%
with a likelihood only $-\log P = 1.1$, indicating that the signal is
consistent with the noise.

A detection of weak coherent shear in the field of 3C336 has already
been suggested by Fort et al.\ (1996) using deep ground-based imaging.
They report a marginal detection with a peak shear signal of $(2.5\pm
1.5)$\% from the analysis of faint galaxy images detected in an 18 ks
$V$ exposure with CFHT in $\sim 0.8$ arcsec seeing.  The  position of
their peak agrees with ours within their $3 \sigma$ error ellipse.  The
peak shear they report is uncorrected for instrumental and atmospheric
degredation, from the simulations in Smail et al.\ (1995a) we would
estimate that the efficiency factor for measuring shear under
conditions similar to those experienced by Fort et al.\ (1996) would be
around $0.4\pm 0.1$.  Correcting their observed shear with this we
obtain an {\it intrinsic} shear strength of $\sim (7\pm 4)$\%, in
reasonable agreement with our value of $(5\pm 1)$\%, after also
correcting for our measurement efficiency.  These independent
detections indicate that a mass over-density exists in the 3C336 field,
which lies in front of a substantial fraction of the faint field
population.  Additional evidence for a possible {\it galaxy}
over-density around 3C336 has also been provided by Hintzen et
al.\ (1991) from their ground-based imaging, where they show that this
field contains a significant excess of galaxies close to the QSO
(Figure~3).  

\begin{figure}
\begin{center}
\psfig{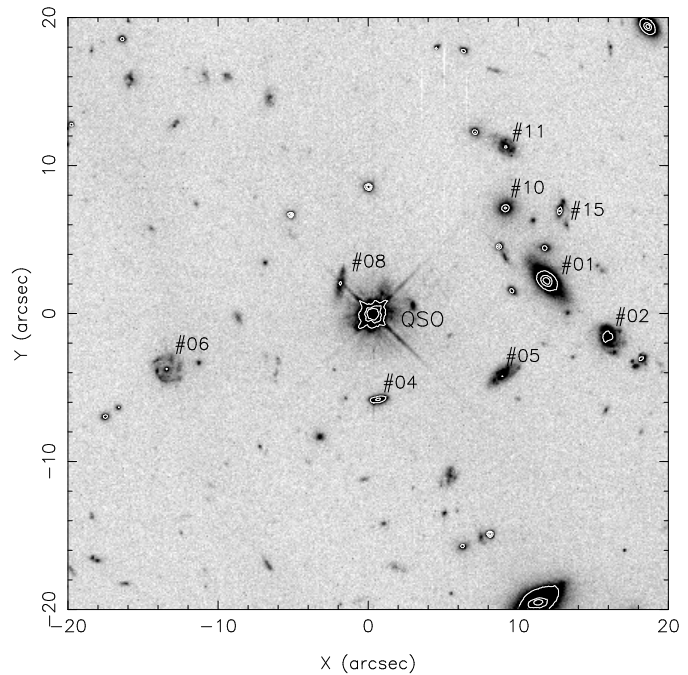}
\caption{A $40'' \times 40''$ area around the radio source 3C336
in F702W from the deeper of our two WFPC-2 exposures of this field. 
We label the various galaxies in this region according to the
naming convention of SD92.  Of particular interest to our study are
the two faint, very red spheroidal galaxies, \# 10 and \#15, which
are proposed to lie at the same distance as the radio source, $z\sim 0.92$.}
\end{center}
\end{figure}

We now discuss other constraints on the possible redshift of the mass
concentration responsible for the shear field we observe.  In this we
are considerably aided by the spectroscopy of galaxies in this field
undertaken by Steidel \& Dickinson (1992, SD92), as well as the
multi-colour imaging of the field from both Hintzen et al.\ (1991) and
SD92.   We start by discussing the morphologies of the galaxies visible
close to the shear peak from the WFPC-2 imaging shown in Figure~3, in
what follows we use the numbering scheme of SD92 to identify the
galaxies.  For a rich cluster we would expect that a significant
proportion of the galaxies associated with the structure would be
early-type systems, consistent with other morphological studies at
similar redshifts (Smail et al.\ 1996b; Faber et al.\ 1996).  Of the
galaxies close to the peak, the most obvious are of course the brighter
and hence typically lower redshift ones (Figure~3).  These include (in
approximately decreasing brightness):  \#1, a bright $z=0.327$ Sa
galaxy; \#2, a face-on spiral with no measured redshift; \# 4, a
late-type spiral at $z=0.472$; \# 5, an Scd at $z=0.661$; \#6, another
face-on spiral with no measured redshift; \#8, a mid-type spiral close
to the QSO for which a tentative detection of [OII]3727 at $z=0.92$ was
reported by Thompson et al.\ (cited by SD92) and \# 11, a small spiral
at $z=0.656$.  As expected most of these bright galaxies are known to
be foreground to the QSO, and all have morphologies typical of the
general field population. SD92 claim that the existence of two galaxies
at $z\sim 0.66$ in this field might indicate a structure at that
redshift,  but if so this system apparently lacks any early-type
galaxies.  Given the lensing mass required for such a structure at
$z\sim 0.6$ (see below), the absence of at least a few early-type
galaxies seems unusual.  Of more relevance are the fainter galaxies,
for which $L\sim L^\ast+1$ at the QSO redshift, these include a large
elliptical (\# 10) and a possible spheroidal galaxy (\# 15).  SD92 show
spectroscopy of \# 10 and state that it is consistent with being an
early-type galaxy at $z\sim 0.92$, further supported by its red
infrared colours.  \# 15 lacks spectroscopic information, but it ranks
alongside \# 10 as being one of the reddest objects in the field,
consistent with an early-type galaxy at $z\sim 0.9$.

We return to discuss the properties of the 3C336 field in Section~5.

\subsection{The Combined Dataset}

Coadding the data-sets from the exposures of the individual fields
allows us to more accurately measure the average shear distortion in
the field and thus to determine, or set an improved limit on, the
typical mass associated with the radio sources. Both 3C336 images have
been excluded from the coaddition, since has already been shown to have
an associated lensing signal. In essence, this test will discriminate
whether the signal seen towards 3C336 is the tip of an iceberg --- with
the other sources being associated with similar mass lenses which we
are missing due to the shallower exposures --- or whether 3C336 (or its
line of sight) is exceptional.

The data taken for GO\# 5401 all have the radio sources in a single
position in the WFPC-2 field-of-view, this allows us to simply merge
the catalogues from the individual fields without having to shift the
field centres.  As long as the radio sources lie within $\sim 300
h^{-1}$ kpc of the cluster centres then the shear fields will add
coherently within our detection aperture.  The merged catalogue
contains 1231 objects, but in order to allow direct comparison with the
results presented above we use the shear detection same algorithm
without attempting to reoptimise the smoothing scale-length.  Despite
the large number of objects in the catalogue, a significant shear
signal is still not detected near the composite source position
(Table~2). The local maxiumum is only slightly more than a 1$\sigma$
deviation, and is associated with an average shear signal of only
0.7\%. The 99\% detection limit that we determine for the merged
catalogue is 1.4\%.   This would indicate that the field around 3C336
is exceptional in the mass concentrations contained within it, or the
position of the radio source relative to them.  

\section{DISCUSSION} 

For many of the individual fields, the shear limits given in Table~2
imply only a very weak contraint on the lensing mass: even at the
present epoch very few clusters have velocity dispersions in excess of
1100~km/s. A tighter limit comes from considering the combined data
set. This is sufficient to set a limit of 700~km/s on the average mass
of the lens under the assumptions made above. If we move the lens to a
lower redshift, or if we are more generous  with the fraction of faint
objects lying at redshifts greater than the radio source (as would be
expected if we used the predicted $N(z)$ from the HDF color modelling),
the limit becomes increasingly stringent (Table~2).  For example,
assuming that $\geq 80$\% of the faint objects lie beyond the radio
galaxy or alternatively moving the lens to $z_{lens}\sim 0.6$ reduces the
allowed velocity dispersion of the `average' structure inhabited by
these luminous radio sources to $\leq 380$~km/s, equivalent to a
central mass of only $M\leq 0.3 \times 10^{14} h^{-1} M_\odot$.

For the deeper 3C336\#2 observation, the detected shear can be
translated into an estimate of the lensing mass.  In the case that the
lens is physically associated with the radio source and roughly 50\% of
the field population lie beyond it, then we obtain $\sigma = 1150\pm
120$ km/s equivalent to a mass of $M_{250} = (2.4\pm 0.5) \times
10^{14} h^{-1} M_\odot$.  Again moving the lens to $z_{lens} \sim 0.6$,
or assuming that $\sim 80$\% of the faint field population to $R\sim
27$ lie beyond $z\sim 1$, brings this estimate down to $\sigma = 670\pm
70$ km/s and $M_{250} \sim 0.8 \times 10^{14} h^{-1} M_\odot$.  The
crucial question for the mass is thus the redshift of the lens.  The
discussion of the 3C336 field given above, while not conclusive, does
indicate that there may be a number of red, spheroidal galaxies in the
region where the gravitational shear peaks.  The limited spectroscopy,
colours and apparent magnitudes are all consistent with these galaxies
lying in a cluster at the radio source redshift.  3C336 has also been
shown to have very extended [OII]3727 emission reaching to $\sim 25
h^{-1}$ kpc (Bremer et al.\ 1992). Bremer et al.\ have modelled the
emission line gas and conclude that the radio source is embedded in a
very large cooling flow, with a mass deposition rate typical of the
most massive local clusters.  There is some support, therefore, for
thinking that the lens is associated with the radio source and thus
represents a massive cluster at $z=0.927$.  Unfortunately for our
attempt to map the $M$--$L_X$ relation at high redshift the QSO 3C336
will contribute to the  X-ray emission, making measurement of the X-ray
luminosity of the underlying cluster emission extremely difficult from
low resolution X-ray imaging.  Nevertheless, the identification of the
shear field in this field, using modest exposures taken from the HST
archive, indicates that an extension of the survey to larger sample may
produce a number of distant cluster candidates which are better suited
to X-ray follow-up. 

If 3C336 does lie in the central regions of a distant massive cluster,
then can we tell if this is true of distant luminous radio sources in
general?  Unfortunately, the relatively low sensitivity of the shear
fields in our other fields means we cannot address this on an
object-by-object basis.  Nevertheless, by combining the 7 fields we can
achieve the required sensitivity.  Analysis of this combined catalog
finds no evidence for a coherent shear around the radio source
position.   The sensitivity of our limit allows us to state that the
majority ($\sim 90$\%) of 3C sources at $z\sim 1$ do not inhabit the
central regions ($\sim 300 h^{-1}$ kpc) of massive clusters ($M_{250}
\geq 0.9 \times 10^{14} h^{-1} M_\odot$, 99\% c.l.).  Given the claim
by Hutchings et al.\ (1995) that the majority of the $z\sim 1$ QSOs
they imaged show strong excesses of faint galaxies within $\sim 100
h^{-1}$ kpc, our observations would limit the masses of the structures
inhabited by these galaxies to $M_{250} \ls 0.9 \times 10^{14} h^{-1}
M_\odot$, consistent with a poor cluster or group.

If instead the mass structure in the 3C336 field is foreground then we
must ask what effect this might have on the observations of the
background radio source and whether amplification bias from such
structures might be a significant feature of the distant radio
catalogues.  Assuming the lens lies at $z\sim 0.6$ and that the shear
peak has a small core radius, we would estimate an amplification of
roughly $\gs 0.3$--1 mag, depending upon the exact size of any core in the
mass distribution.  Bartelmann \& Schneider (1992) have simulated the
expected amplification for random lines of sight.  Taking our typical
shear sensitivity, $\sim 3$\%, their simulations indicate that roughly
13\% of sightlines would show shear peaks above this value.  More
recent work by Wambsganss et al.\ (1996) would indicate a similar
fraction.  It should also be noted that if amplification bias is a
strong effect in the 3C catalog then we should obtain detections in a
higher proportion of fields than the `random' expectation, perhaps even
doubling the frequency.  This range would then consistent with our
observation of $\sim 10$--20\% (1--2 of 8, depending on the detection
aperture used) although significantly more data would be required to
conclusively test this prediction.   

\section{CONCLUSIONS}

We now restate our main conclusions:

\noindent{$\bullet$} We have searched the fields of 8 $z\sim 1$
luminous radio sources, both radio galaxies and radio-loud quasars for the
signature of gravitational lensing.  To achieve this we have analysed the
coherence of the shear field determined from the shapes of faint
galaxies ($R\sim 26$) in deep archival HST/WFPC-2 images.

\noindent{$\bullet$} We find evidence for a significant shear
field ($5\pm 1$\%) peaked close to the radio-loud QSO 3C336 at $z=0.927$.
The mass concentration reponsible is present in two independent
WFPC-2 images, as well is a previously reported ground-based
observation by Fort et al.\ (1996).  

\noindent{$\bullet$} The characteristics, magnitudes and colours, of
the early-type galaxies lying close to the shear peak are consistent
with them lying at high redshift ($z\sim 1$), as is the limited
spectroscopic information.  In addition the coherent shear signal is
not detected in a brighter sample of galaxies in the field which can
reasonably be expected to lie at $z\ls 1$.  We therefore propose that
the shear field may be produced by a cluster associated with the
radio source at $z=0.927$.  In this case, adopting a redshift
distribution for the faint field galaxies in our analysis ($<\! z\! >
\sim 1$), we infer a mass of $M_{250} \sim 2.4 \times 10^{14} h^{-1}
M_\odot$ within $250 h^{-1}$ kpc for the cluster.  Placing the field
population at higher redshifts, or moving the lens in front of
the radio source will decrease this mass estimate.

\noindent{$\bullet$} In the remaining 7 fields analysed we find no
evidence for coherent shear patterns around the radio sources.
However, the individual exposures are sufficiently shallow that this
conclusion is not strong, the 99\% confidence limit on the mean mass is
$M_{250}\leq 2.7 \times 10^{14}h^{-1} M_\odot$.  In an attempt to
strengthen it we combined the shear fields in the 7 exposures to obtain
a higher sensitivity in our shear measurement.  Using this approach we
are able to set an upper limit on the mean induced shear associated
with any mass structures in these 7 fields corresponding to
$M_{250}\leq 0.9 \times 10^{14}h^{-1} M_\odot$.  This implies that the
average $z\sim 1$ luminous radio source does not typically inhabit the
central regions of massive clusters. 

\noindent{$\bullet$} The absence of a detected shear pattern in the
composite catalogue also indicates that the bulk of these distant radio
sources are not strongly lensed by massive foreground structures
(i.e.\ considerably larger than individual galaxy haloes).  Our
observations appear consistent with the expectations for lensing of
distant radio sources within standard cosmological models
(e.g.\ Bartelmann \& Schneider 1992; Wambsganss et al.\ 1996) although
significantly larger samples would be needed to test these predictions.

\noindent{$\bullet$} We have demonstrated that it is easily within
the capabilities of HST to detect shear fields associated with moderate
mass $z\sim 1$ clusters in modest integrations ($\ls 10$ orbits). 
Expanding these observations to a larger sample of candidates
will be a necessary first step to understand the X-ray observations
of distant clusters which will come from the next generation of
X-ray telescopes.  These HST observations will also greatly benefit
from the larger field of view available from the Advanced Camera
when it is added to the HST instrument complement in 1999.

\section*{ACKNOWLEDGMENTS}

RGB and IRS acknowledge support from the University of Durham
and computer resources supported by the Starlink project.
IRS also acknowledges support through a PPARC Advanced Fellowship.
We would like to thank Ale Terlevich and Jean-Paul Kneib for their
helpful comments and assistance.

\vfil\eject

\begin{table*}
\caption{Log of observations. 
The columns give: (1) the target name;
(2) coordinates; (3) radio source redshift;
(4) classification for the radio source
(RG $\equiv$ radio galaxy, RLQ $\equiv$ radio-loud quasar); (5) exposure
time; (6) filter; (7) 50\% completeness limit of the faint galaxy catalogue;
(8) number of galaxies in the final sample and (9) the proposal ID.}
\vspace{0.5cm}
\begin{tabular}{lccclrccrc}
\noalign{\medskip}
\hline\hline
\noalign{\smallskip}
{ID} &  {RA } &   {Dec } & $z$ & Type & T$_{\rm exp}$ & Filter &  $R_{max}$ & $N$ 
& Proposal \cr
     & (J2000) & (J2000) & (ks) & & & (50\%) & & \cr
\hline
\noalign{\smallskip}
3C2   & 00 06 25.3 & $-$00 03 59 & 1.037 & RLQ & 17.6 & F675W & 26.1 &  157 &  GO-5401 \cr
3C212 & 08 58 38.2 & $+$14 09 57 & 1.048 & RLQ & 17.6 & F675W & 26.1 &  167 &  GO-5401 \cr
3C217 & 09 08 46.7 & $+$37 48 30 & 0.897 & RG  &  7.2 & F622W & 25.6 &   97 &  GO-5401 \cr
3C237 & 10 07 56.9 & $+$07 30 31 & 0.877 & RG  &  8.0 & F622W & 25.6 &  111 &  GO-5401 \cr
3C245 & 10 42 41.5 & $+$12 03 50 & 1.029 & RG  & 17.6 & F675W & 26.1 &  133 &  GO-5401 \cr
3C280 & 12 57 01.8 & $+$47 20 48 & 0.996 & RG  &  8.8 & F622W & 25.6 &  108 &  GO-5401 \cr
3C289 & 13 45 29.9 & $+$49 47 05 & 0.967 & RG  &  8.0 & F622W & 25.6 &  143 &  GO-5401 \cr
3C336 & 16 24 37.4 & $+$23 45 56 & 0.927 & RLQ &  7.2 & F622W & 25.6 &  106 &  GO-5401 \cr
3C336\#2 &         &             &       &     & 24.0 & F702W & 27.0 &  330 &  GO-5304 \cr
\noalign{\smallskip}
\noalign{\hrule}
\noalign{\smallskip}
\end{tabular}
\end{table*}

\begin{table*}
\caption{Results of Shear Pattern Analysis.  Here we list: (1) the field
name; (2) peak shear detected within a 200 $h^{-1}$ kpc aperture centered
on the radio source; (3) 99\% confidence limit on the upper bound of the
shear and (4) the probability of observing the peak shear value estimated
from Monte Carlo simulation.  The final two columns give the limits on
the velocity dispersion of a singular isothermal sphere model for the
lensing mass given the observed shear, or the limits on it.  The two
columns give the estimates depending upon whether the lens is placed at
the radio source redshift (5, $z_{gal}$) or in the foreground (6) at
$z_{lens}=0.6$.  The latter also gives some indication of the velocity
dispersion expected if the lens lies at the radio source redshift, but
a larger fraction of the field population lies beyond it than expected
from our adopted $N(z)$. }
\vspace{0.5cm}
\begin{tabular}{lccccc}
\noalign{\medskip}
\hline\hline
\noalign{\smallskip}
{ID} &  shear & 99\% det. & Likelihood & $\sigma(\zl =\zr)$ & $\sigma(\zl = 0.6)$ \cr 
     &    (\%) &      limit &  ($-\log P$) &  (km/s) & (km/s) \cr
\hline
\noalign{\smallskip}
3C2   &  1.9 &  3.0 & 1.20 &    $<1100$ &  $<540$\cr
3C212 &  2.3 &  3.2 & 1.27 &    $<1240$ &  $<590$\cr
3C217 &  0.0 &  4.2 & 0.42 &    $<1310$ &  $<750$\cr
3C237 &  1.2 &  3.7 & 0.68 &    $<1080$ &  $<670$\cr
3C245 &  2.5 &  3.3 & 1.42 &    $<1100$ &  $<550$\cr
3C280 &  1.7 &  4.4 & 0.79 &    $<1470$ &  $<680$\cr
3C289 &  3.5 &  3.3 & 1.95 &    $<1230$ &  $<640$\cr
3C336 &  3.9 &  3.8 & 2.15 &    \cr
3C336\#2 &  4.0 &  2.5 & 3.30 &  $1150\pm 120$ &  $670\pm 70$\cr
Combined$^a$ 
      & 1.25&    0.7 &  1.40 &   $<710$ &  $<380$\cr 
\noalign{\smallskip}
\noalign{\hrule}
\noalign{\smallskip}
\end{tabular}
\bigskip

\begin{tabular}{cl}
(a) & Combined field consists of all images excluding 3C336 and 3C336\#2.
The upper mass limit is computed using\cr 
~ & the average radio source redshift of
$z_{\rm gal}= 0.979$ and a mean background field redshift of $<\! z\! > = 1.05$.\cr
\end{tabular}
\end{table*}

\end{document}